\documentclass[twocolumn,showpacs,preprintnumbers,amsmath,amssymb]{revtex4}
\usepackage{graphicx}
\usepackage{dcolumn}
\begin{document}
\title{Importance of convection in the compaction mechanisms 
of anisotropic granular media}
\author{Philippe Ribi\`ere, Patrick Richard, Renaud Delannay and Daniel Bideau}
\affiliation{Groupe Mati\`ere Condens\'ee et Mat\'eriaux, UMR CNRS 6626, Universit\'e de Rennes 1, Campus de Beaulieu, F-35042 Rennes cedex}
\date{\today}
\begin{abstract}
We report the experimental observation of novel vortex patterns
in a vertically tapped granular media.
Depending on the tapping acceleration 
two behaviors are observed.
For high acceleration 
a convection vortex appears
in the whole media whereas for low acceleration two unstable vortices
appear in the upper part of the media and slowly compact the lower part.
We explain the formation of the vortices and relate them to 
granular convection. Our results demonstrate
the importance  of compression waves propagation on granular compaction.

\end{abstract}
\maketitle
In hydrodynamics, flow instabilities lead to the formation
of vortex patterns which affect the behavior of the fluid.
It is well known that granular media can exhibit both fluid-like 
and a 
solid-like behavior and thus 
do not behave like classical fluids.
In order to understand the specificities of granular
media, some studies of instabilities have been 
carried out for vertical vibrated granular 
layers~\cite{Evesque1989,Umbanhowar1996,Blair2003},
granular flows~\cite{Louge2001,Pouliquen1997,Forterre2001} and gazes~\cite{Falcon1999a}.
Moreover, a granular packing under vertical shaking
leads to a global 
compaction~\cite{Chicago,Philippe2002,Richard2003} and this 
compaction can be linked to 
convection~\cite{Philippe2002,Philippe2003,Ribiere2004}.
Since convection in fluids can lead to instabilities,
can granular systems under compaction exhibit vortices ?
In this letter we present a vortex creation process that occurs
in a tapped 
pile of anisotropic grains
and explain 
the mechanisms of such a creation.
Since it has been recently proved that grain anisotropy
is important in pattern formation~\cite{Blair2003}
we focus on anisotropic granular materials.

%
%
The experimental setup consists of a glass cylinder of 
diameter $ D \approx 10\mbox{ cm}$ filled with 600~g of grains
(corresponding to a height of roughly $10\mbox{ cm}$).
This container is placed
on a plate connected to an electromagnetic exciter (LDS V406) which
induces a vertical displacement of the plate. The container is, in this
way, regularly shaken ($\Delta t = 1 $s)
by vertical taps. Each tap is
created by one entire period of sine wave at a fixed frequency $f= 30$~Hz.
The resulting motion of the whole system, monitored by an accelerometer
at the bottom of the container, is however more complicated than a
simple sine wave. At first, the system undergoes a positive acceleration
followed by a negative peak with a minimum, $-\gamma_{max}$.
After the
applied voltage stops, the system relaxes to its normal repose position.
This negative peak acceleration is used to characterize the tap
intensity by the dimensionless acceleration $\Gamma = \gamma_{max}/g$
(with $g = 9.81\mbox{ m.s}^{-2}$).
It should be pointed out that we only used
taps for which the grains took off from the bottom
of the glass cylinder i.e. above the lift-off threshold.
%
%
The packing fraction is measured using a $\gamma$-ray 
absorption setup~\cite{Philippe2002}.
The results presented here have been obtained using 
long rice
(basmati rice).
In order to quantify the anisotropy, each grain is assimilated
to an ellipsoid.
The mean ratio between the two axes is  
$2.5$ with $10\%$ of dispersion. The length of the longer axe
will be noted $d$.
The method used to build our initial packing, already used for 
spheres~\cite{Philippe2002}, is reproducible
(we obtain the same initial packing fraction $\Phi\approx 0.56$
and the same
packing fraction evolution for a given $\Gamma$).
Sequences of $10^4$ to $10^6$ taps are carried out.
The only control parameter is the tapping intensity $\Gamma$.
Here we call "time" the number of taps and the
"dynamics" is the succession of static equilibrium induced
by the taps.
{The mean packing fraction increases monotically with time following
stretched exponential. All the details on the compaction
dynamics can be found elsewhere~\cite{Ribiere2004}.}

\begin{figure}[htbp]
\includegraphics*[width=7cm]{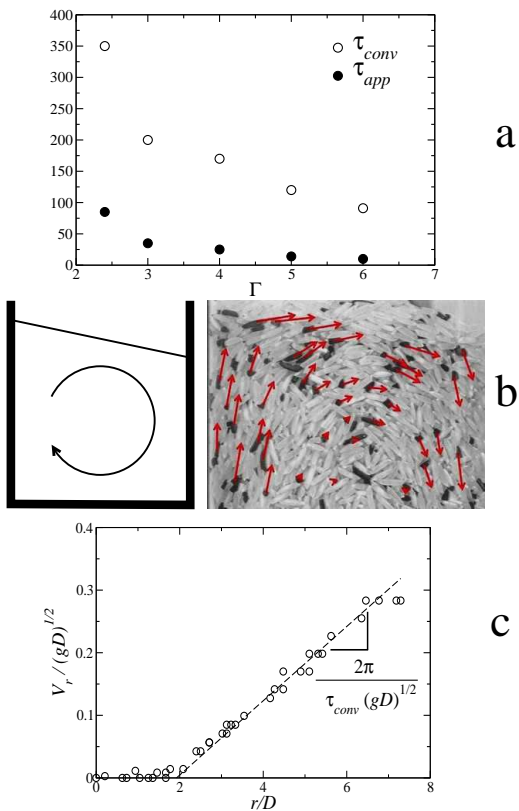}
\caption{
a) $\tau_{app}$ and $\tau_{conv}$ (see text for definition)
versus tapping intensity $\Gamma$.
b) Sketch of the convection obtained for 
basmati rice and for
$\Gamma = 6$ (left) and snapshot of the medium during 
relaxation (right) where convection roll can be clearly
seen on the walls.
c) Radial displacement profile : the granular medium
rotate like a solid around an immobile core.}
\label{fig:vortex}
\end{figure}

\begin{figure}[htbp]
\includegraphics*[width=7cm]{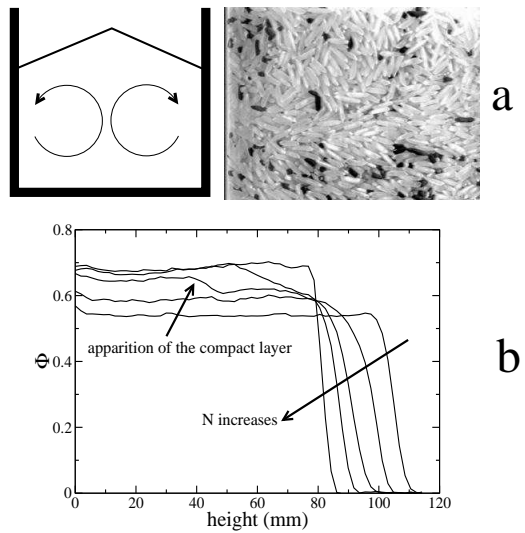}
\caption{
a) Sketch of the convection obtained for
basmati rice and for
$\Gamma = 2.4$ (left) and snapshot of the medium during
relaxation (right). The compact zone can be observed on the side walls.
b)Packing fraction profile obtained for basmati rice at $\Gamma = 2.4$
for various number of taps (see the arrow) : $N$ = 0, 10, 100, 3116, 10000, 31162 and 100000.
}
\label{fig:profil}
\end{figure}

The first observation is that convection takes place in
the medium : after $\tau_{app}$, a given time depending
on $\Gamma$, one or two counter-rotating vertical convection rolls
appear. 
{They are easily identified by tracking grains position
after each tap.}
To characterize them, we measure the time
it takes a grain to revolve around the center of the roll, $\tau_{conv}$.
Fig.~\ref{fig:vortex}a reports
the variations of $\tau_{conv}$ and $\tau_{app}$ with
$\Gamma$. As expected, the greater the tapping acceleration is,
the smaller these two times are.
Concerning the number of rolls, two cases
should be considered.
For high tapping intensity (typically $\Gamma > 3$) convection
takes place
within the whole medium. After about ten taps, two convection rolls
appear but this situation becomes unstable.
One of the rolls progressively disappears and after this transient the
whole medium is occupied by only one roll and
the free surface of the medium is tilted from the horizontal
(for example about $20^\circ$ for $\Gamma=6$).
%
%
A sketch and a snapshot of this convection roll is
reported Fig~\ref{fig:vortex}b. 
{The arrows represent in arbitrary units the displacement field of the grains
and clearly demonstrate the existence of convection.
A radial displacement profile is reported Fig~\ref{fig:vortex}c.
An immobile core 
can be observed. In an ideal system this non-mobile part
should be reduced to a point,
but due to the formation of spatially-correlated clusters
the size obtained is around 4 $d$. After this part a linear
profile is found with a slope compatible with the value
of $\tau_{conv}$. This part of the packing rotates around 
the immobile core like a solid.
}
For lower values of the tapping intensity 
(typically $\Gamma < 3$) two convection rolls 
roughly equally-sized  and
localized near the free surface 
appear.
The other part of this packing (below
the rolls) is not submitted to convection.
As for the high tapping intensity case, this
situation is unstable. Contrary to the previous
case the two rolls do not merge : their size
continuously decreases
until they totally disappear. 
Note that the convection turns out not to be a
phenomenon in competition with
compaction. 
On the contrary the convection rolls create at their base a 
band of high packing fraction.
On Fig.~\ref{fig:profil}a taken for
$\Gamma=2.4$ after 1500 taps,
this band is clearly visible at the base of the two
convection rolls 
and one can observe that the grains in
this band have a preferential
horizontal orientation, due to the motion of the rice in convection.
The band of high density can be distinguished
on the packing fraction profile (Fig~\ref{fig:profil}b).
The convection rolls create the high density
zone and their size decrease corresponds to
the growth of this zone.
The orientation of the grains 
is totally different
from the one observed in another compaction experiment~\cite{Villarruel2000}.
Indeed, the vessel used in~\cite{Villarruel2000} is very narrow (1 grain length)
compare to our (about 20 grain lengths). Thus the nematic ordering observed
in this tube is likely due to the strong confinement
and no convection
is possible~\cite{Blair2003}. Note that the aspect ratio of the grains
is probably also an important parameter. 

\begin{figure}[htbp]
\includegraphics*[width=7cm]{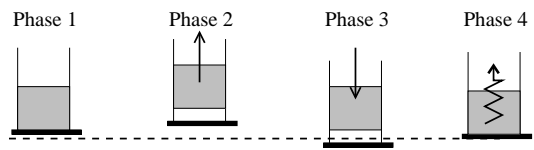}
\caption{Sketch  of the four phases of a tap.
The dashed line corresponds to the repose position of the vessel.
The medium follow the motion of the vessel (phase 1),
takes off from the bottom (phase 2), lands
on the bottom (phase 3) and finally the
compression wave propagates (phase 4).}\label{fig:phases}
\end{figure}

\begin{figure*}[htbp]
\includegraphics*[width=14cm]{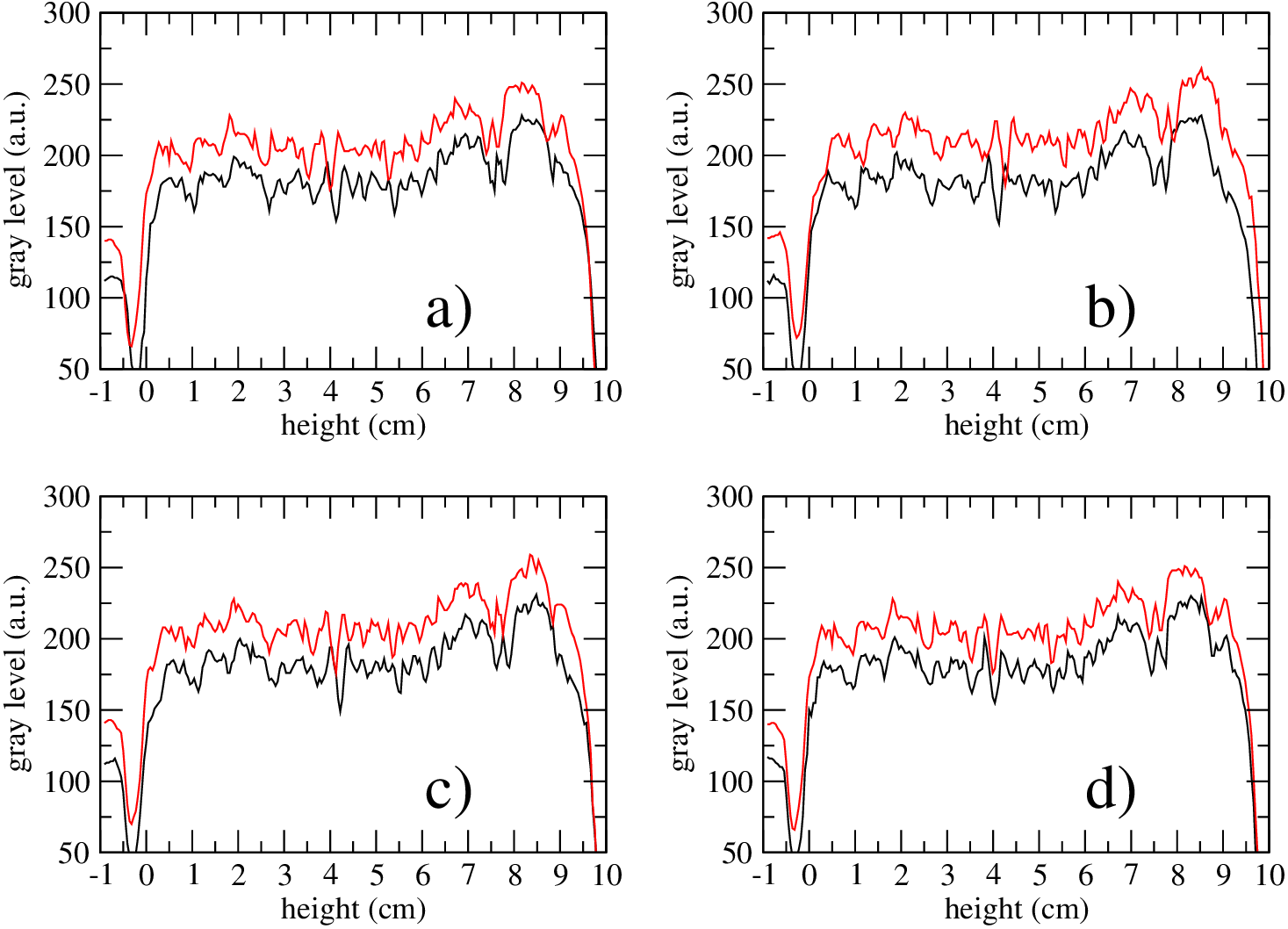}
\caption{Gray level profiles along a vertical line
on the picture of the system for $\Gamma = 2.4$ : 
a) 
before (down) and after the first tap (up),
b)
just before the medium take off (down) and at its
higher position (up),
c) at its higher position (down) and just after the medium lands  (up),
d) just before (down) and after (up) the compression wave.
For clarity, each curve has an offset in the y-axis
and is reported in the reference frame of the vessel.}
\label{fig:gray2.4}
\end{figure*}

To understand the convection creation and evolution
we have used a high-speed camera ($1000$ images per second)
to analyze the motion and the deformation of the packing 
during  a tap. Four phases can be 
visually distinguished (see Fig.~\ref{fig:phases}): 
\begin{itemize}
\item the rice follows the motion of the plate (phase 1)
\item the rice takes off from the bottom (phase 2)
\item the rice lands on the bottom (phase 3)
\item a compression wave propagates throughout the medium (phase 4)
\end{itemize}
The rate of compaction and dilatation is measured during each 
phase for the first tap 
 (the motion of
the grains is greater here than during the other solicitations
and
thus the detection of motion is easier.
The data were averaged on two different experiments
and 
no significant differences were observed.
To track the grains, we use an image processing software
that computes the gray level profile along a vertical
line at the side wall. 
For clarity each profile is reported in the reference frame of
the vessel.
\begin{figure*}[htbp]
\includegraphics*[width=14cm]{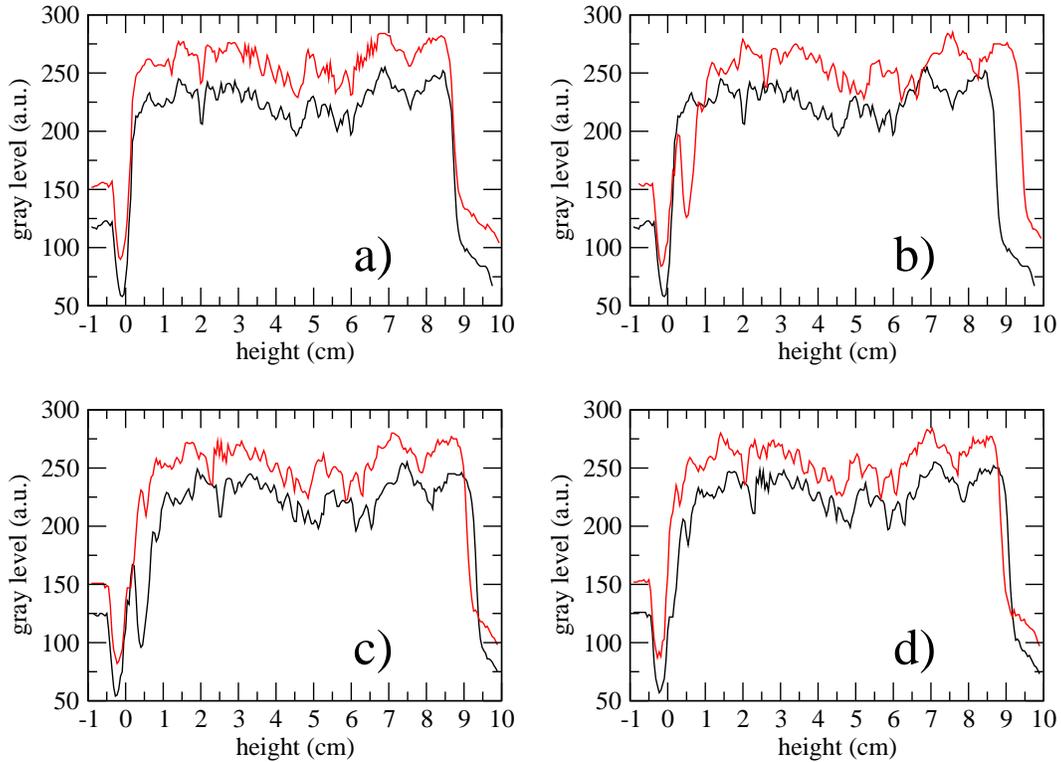}
\caption{Gray level profiles along a vertical line
on the picture of the system for $\Gamma = 6$ : 
a) 
before (down) and after (up) the first tap,
b)
just before the medium takes off (down) and at its
higher position,
c) at its higher position (down) and just after the medium lands (up),
d) just before (down) and after the compression wave.
For clarity, each curve has an offset in the y-axis
and is reported in the reference frame of the vessel.}
\label{fig:gray6}
\end{figure*}
To improve contrast and thus to have sharp peaks on the gray level
profile, 8\% of the grains are paint in black.
We can then follow the displacement of the peaks between
each picture {(1000 pictures per seconds)
and thus detect dilatation, compaction
or wave propagation during each phase of the motion}. The precision on the peak displacement
is $0.3$ mm for a packing height of about $10$ cm.
Note that, 
in order to avoid overlapping, each profile presented here
is shifted vertically.
On Fig~\ref{fig:gray6}a,
the two gray level profiles correspond respectively to the
state before and after the first tap of $\Gamma=6$. 
These two profiles
 are similar and the peaks present 
can be easily tracked.
Let us now examine with this method the phases of the motion.
During the first phase, the rice follows the motion of the container.
As the medium is in a
stable configuration, 
this motion does not create a relative displacement of  the grains.
As the cylinder slows down, the medium takes off from
the bottom and a dilatation of
the piling is observed (Fig~\ref{fig:gray6}b).
This dilatation is not uniform : $1$ mm for the 
upper part of the packing and $0.5$ mm for the lower part.
A study of the orientation of the grains during this phase shows
that they have a tendency to align vertically.
Moreover, as already observed for glass beads in~\cite{Philippe2003},
all the packing takes off and its bottom is still 
parallel to the plate.
During the third phase (Fig~\ref{fig:gray6}c), the rice lands,
which induces a compaction of the medium.
Surprisingly this compaction  
is smaller than the dilatation monitored during the
second phase and at 
this stage the medium is less
compact than initially.
Note that the vertical orientation tendency of the rice always
exists even if it is less visible than during the second phase.
In fact, the medium mainly compacts during the last phase
(propagation of the compression wave).
The evolution of the gray level during that phase is 
reported on Fig~\ref{fig:gray6}d and compaction of the
medium is indeed observed. {
The compression wave can be easily tracked on the fast camera
movie. Its velocity can then be measured. For $\Gamma=6$ 
we have $V\approx16.5\mbox{ m.s}^{-1}$.}
Moreover, it is only during this phase of
propagation that  the contribution to convection appears.
Indeed, during phases 1, 2 and 3 the bottom of the packing
remains parallel to the plate, hence no
convection-like grain motions are observed. 
This result clearly demonstrates that 
compaction and convection are strongly linked.
{The compression wave is a consequence of the dissipation
of the interaction energy stored by the grains
during the previous motion phases. This allows 
the grains to move slightly around their initial
position and to possibly find a more stable position
corresponding to 
an higher value of the packing fraction.}
{The velocity of the compression wave is
of the same order of magnitude than those obtain in the same
setup  with glass beads \cite{Philippe2003} %
but smaller than the
values obtained in a one dimensional granular medium~\cite{Falcon1997}}.\\
{It should be pointed out that 
in our set-up compaction, is mainly associated to 
compression wave propagation and convection.
Nevertheless it is not the only one, since compaction
can be observed even without convection~\cite{Chicago}.
This also explain why the characteristic times found in~\cite{Philippe2002}
are smaller than those obtained in~\cite{Chicago}.}\\
For $\Gamma=6$, the effect of the 
compression wave is easy to analyze for two reasons. 
First, the dilatation of the medium just before the wave is still 
important and its effect  
is easier to appreciate.
Second, the intensity of the compression wave is 
high because the medium lands on the plate
when the plate comes up again.\\
We have also performed such an analysis for 
$\Gamma = 2.4$.
On 
Fig~\ref{fig:gray2.4}a, the two gray level profiles
reported 
correspond respectively to the
state before and after the first tap.
The motion of the grains is smaller
than for $\Gamma=6$. 
Many characteristics of the medium motion
are similar to those obtained for $\Gamma=6$.
During the first phase (the grains
follow the plate motion)
the medium is not
affected. 
During the second phase, we monitor a small dilatation in the medium.
Nevertheless
the mean orientation does not change as much as for 
$\Gamma = 6$.
During the third phase, the packing fraction increases.
What is surprising and different from the
$\Gamma=6$ case is that during the last
phase, the compression wave propagates inhomogeneously through
the medium. Indeed
no
motion is visible at the bottom whereas a small displacement
of $1$ mm is monitored
on the
top of the packing.
This is in agreement with previous analysis : 
the wave does not affect the bottom part of the packing
where no convection is observed, contrary to the upper
part where both convection and compression wave propagation are present.
Note that, as expected, the effect of the compression wave is less important
for $\Gamma=2.4$ than for $\Gamma=6$. \\
We explain this behavior difference by the motion
of the packing relatively to the bottom plate.
For $\Gamma=2.4$, the medium lands as the plate is moving downwards.
Therefore, the apparent fall velocity 
of the medium is smaller than in motionless case and the intensity
of the compression
wave is smaller than for $\Gamma=6$.
The other observation is that there is also a 
strong correlation between the 
apparition of convection and of the compression wave. Indeed, in all our
experiments the compression wave
creates a motion only in the part of the medium where
convection will take
place during the other solicitations. 
This explanation helps understanding the transition between the
two behaviors :
for $\Gamma=3$, the fall of the medium coincides with the motionless, lowest position of the plate.

To summarize, the new instability presented in this paper leads to a
vortices formation during compaction of anisotropic granular media
at relatively low tapping intensity.
Below the vortices, an ordered compact zone can be created where
the grains are oriented horizontally. 
By analyzing the packing behavior during one tap
we  showed that the main compaction mechanism is due to a compression wave
propagation. 
Furthermore this compression wave propagation creates convection
proving the strong correlation between granular compaction and convection.


\begin{thebibliography}{0}
\bibitem{Evesque1989} P. Evesque and J. Rajchenbach, Phys. Rev. Lett., {\bf 62}, 44 (1989).
\bibitem{Umbanhowar1996}
P. B. Umbmbanhowar, F. Melo and H. L. Swinney, Nature 382(6594), 793 (1996).
\bibitem{Blair2003} D. L. Blair, T. Neicu and A. Kudrolli, Phys. Rev. E {\bf 67}, 031303 (2003).
\bibitem{Louge2001}
M. Y. Louge and S. C. Keast, 
Phys. of Fluids {\bf 13(5)},  1213
(2001).
\bibitem{Pouliquen1997}
O. Pouliquen, J. Delour and S.B. Savage, Nature (London) {\bf 386}, 816 (1997)
\bibitem{Forterre2001}
Y. Forterre and O. Pouliquen,
 Phys. Rev. Lett. {\bf 86}, 5886 (2001).
\bibitem{Falcon1999a}
E. Falcon., R. Wunenburger, P. Evesque, S.  Fauve, C. Chabot., Y. Garrabos,  and D. Beysens,  
           Phys. Rev. Lett., {\bf 83}, 440(1999). 
\bibitem{Falcon1997} E. Falcon, PhD thesis, Lyon (1997).
\bibitem{Chicago}
    J.B. Knight, C.G. Fandrich, C.N. Lau, H.M. Jaeger and
    S.R. Nagel, Phys. Rev. E, {\bf51}, 3957 (1995);
    E.R Nowak, J.B. Knight, E. Ben-Naim, H.M. Jaeger and
    S.R. Nagel, Phys. Rev. E {\bf57}, 1971 (1998).
\bibitem{Philippe2002}
    P. Philippe and D. Bideau, Europhys. Letters, {\bf60}, 677 (2002).
\bibitem{Richard2003}
P. Richard, P. Philippe, F. Barbe, S. Bourl\`es, X. Thibault and D. Bideau,
Phys. Rev. E, {\bf 68}, 020301(R) (2003).
\bibitem{Ribiere2004}
    P. Ribi\`ere, P. Richard, D. Bideau, R. Delannay in preparation.
\bibitem{Philippe2003} 
  P. Philippe and D. Bideau, Phys. Rev. Lett. {\bf 91} 104302 (2003).
\bibitem{Villarruel2000}
F. X. Villarruel and B. E. Lauderdale and D. M. Mueth and H. M. Jaeger
Phys. Rev. E. {\bf 61}, 6914 (2000).
%
%













 
 
  


        



\end{thebibliography}
\end{document}